\documentclass[twocolumn, tighten, twocolappendix]{aastex63}

\usepackage{amsmath}
\usepackage[flushleft]{threeparttable}

\shorttitle{Flat $\Lambda$CDM Probe of BAOl}
\shortauthors{Jayson}

\begin{document}

\title{Probing BAO studies with flat $\Lambda$CDM constraints}

\email{1jsjayson@gmail.com}

\author{Joel S. Jayson}
\affil{P.O. Box 34 \\
Brooklyn, NY 11235 USA}

\begin{abstract}

One of the algorithms typically used for fitting anisotropic Baryon Acoustic Oscillations (BAO), $\alpha_{\perp}D_{M, \text{fid}}(z)/r_{d, \text{fid}}=D_{M}(z)/r_{d}$, provides an excellent fit as $z\Rightarrow 0$.  Yet, at $ z_{\text{drag}} \simeq 1060$, $D_{M, \text{fid}}(1060)/r_{d, \text{fid}}\simeq D_{M}(1060)/r_{d}$. Since $\alpha_{\perp}$ does not change with increasing redshift, we conclude that the algorithms deviate from true solutions with increasing $z$.  We investigate that departure through an assumption of a flat $\Lambda$CDM cosmology.  At the relatively low redshifts at which isotropic BAO studies were performed, the divergence is smaller than study uncertainties. However, while the transverse measure, $\alpha_{\perp}$ provides a reliable value in real space, not corrupted by redshift distortion,  the isotropic measure, $\alpha$, has a radial component, and  is influenced by redshift distortion. That distortion recommends against the inclusion of isotropic studies in cosmological parameter evaluations. At the high redshifts of Ly$\alpha$ forest BAO studies, $z\sim$ 2.35, the deviations of the anisotropic BAO algorithms from the flat $\Lambda$CDM solutions  give rise to markedly deceptive results. We replicate an example, in which those algorithms lead to a value of  $\Omega_{m}$ which is more than 15 per cent lower than that of the flat  $\Lambda$CDM cosmology computation. The BAO algorithms are inappropriate for the next generation of BAO studies.  A flat $\Lambda$CDM analysis offers an alternative means of assessment.  

\end{abstract}

\keywords{large scale structure of the universe--- 
cosmic microwave background radiation --- cosmological parameters --- early Universe}

\section{Introduction} \label{sec:intro}

 Baryon Acoustic Oscillation (BAO) findings taken together with cosmic microwave background (CMB) results significantly restrict  the $\Omega_m$, $H_0$, confidence interval.  CMB satellite missions have applied that joint characteristic \citep{Spergel2007,Komatsu2011,Hinshaw2013,Ade2013,Ade2015,Aghanim2018} since the first successful statistical recoveries \citep{Eisenstein2005,Cole2005} of the BAO acoustic scale.  \citet{Addison2013,Addison2018} performed analyses of ensembles of several BAO studies, constraining cosmological parameters, and probing the CMB versus low redshift measurement discrepancy of the Hubble constant, and \citet{Aubourg2015} have employed a set of BAO studies to survey variations from a flat, $w$=-1, $\Lambda$CDM universe.

  Evaluations of the BAO feature from the two point correlation function (2PCF) have attained a precision of $\sim$1 percent, (see, e.g, \citet{Cuesta2016}). Projects recently initiated, Dark Energy Spectroscopic Instrument (\textit{DESI}) \citep{Vargas2019}, and further in the future, \textit{Euclid}, \citep{Euclid2019}, and  the Wide Field Infra Red Space Telescope (\textit{WFIRST}) \citep{Dore2019} will attain sub-percent precision levels.  With that improved precision in mind, we explore BAO methodology with the intent of identifying uncertainties that might limit attainable precision.  
  
At this juncture, precise CMB measurements, together with gravitational lensing, have restricted variation from flatness to $\Omega_{K}=-0.012^{+0.021}_{-0.023}$, \citep{Simard2018}, and combined CMB and SNe~Ia measurements constrain $w$ to $-1.02\pm{0.06}$ \citep{Betoule2014}.  We rely on these restrictions to justify an analysis of BAO results, in which we set $\Omega_{K}=0$ and $w=-1$.   That flat $\Lambda$CDM model scaffolding facilitates the derivation of cosmological parameter values from the BAO studies.  \citet{Percival2002} derived an expression, $\Omega_{m}h^3$=constant, which describes the $\Omega_m$, $H_0$ near-degeneracy within the flat cosmology, where the dimensionless parameter, \textit{h}, is used to define the Hubble constant, $H_{0}=100h$ km $\text{s}^{-1}$ $\text{Mpc}^{-1}$. The \textit{Planck} mission provides a premium data set, from which  we extract their value of the constant \citep{Aghanim2018},
\begin{equation}\label{E:degen}
 \Omega_{m}h^3=0.09633\pm 0.00029.
 \end{equation}

 The standard deviation is low enough, 0.3 per cent, to enable us to work with the mean value alone throughout this paper with no unfavorable effect. In the following section we use that expression to derive a relationship between the BAO 2PCF feature location and values of $\Omega_m$ and $h$.  We compute matter power spectra to locate the BAO feature, and towards that end, require a complete parameter set.  In addition to $\Omega_m$ and $h$, that parameter set is comprised of the baryon mass fraction, $\Omega_{b}h^2$, the spectral index, $n_s$, and the reionization optical depth, $\tau$. We follow the lead of \citet{Addison2018} in using an empirical value of $\Omega_{b}h^2$ based on measurements of the primordial deuterium abundance.  Our value, $\Omega_{b}h^2=0.0224$, is derived from \cite{Cooke2018}.  The scalar spectral index, $n_s$  is set at 0.965, an average of values found in several CMB parameter determinations, \citep{Calabrese2017,Ade2015,Hinshaw2013}, and the reionization optical depth  is assigned a value of, $\tau$=0.054, as per \citet{Aghanim2018}. 

In Section~\ref{sec:BAOsol}, we calculate the 2PCF over a range of flat $\Lambda$CDM cosmologies, and derive simple relationships between $\Omega_m$, $h$, the fiducial peak position of the BAO feature, and $\alpha$, the ratio of the fiducial peak position to the measured peak position.   Section~\ref{sec:anal} applies the analysis to isotropic and anisotropic BAO studies, and investigates the applicability of the commonly used BAO algorithms.   Section~\ref{sec:Conclusions} sets forth our conclusions. 

 \section{Test Bed}
 \label{sec:BAOsol}
 
  \begin{figure} 
\centering
 \includegraphics[width=\columnwidth]{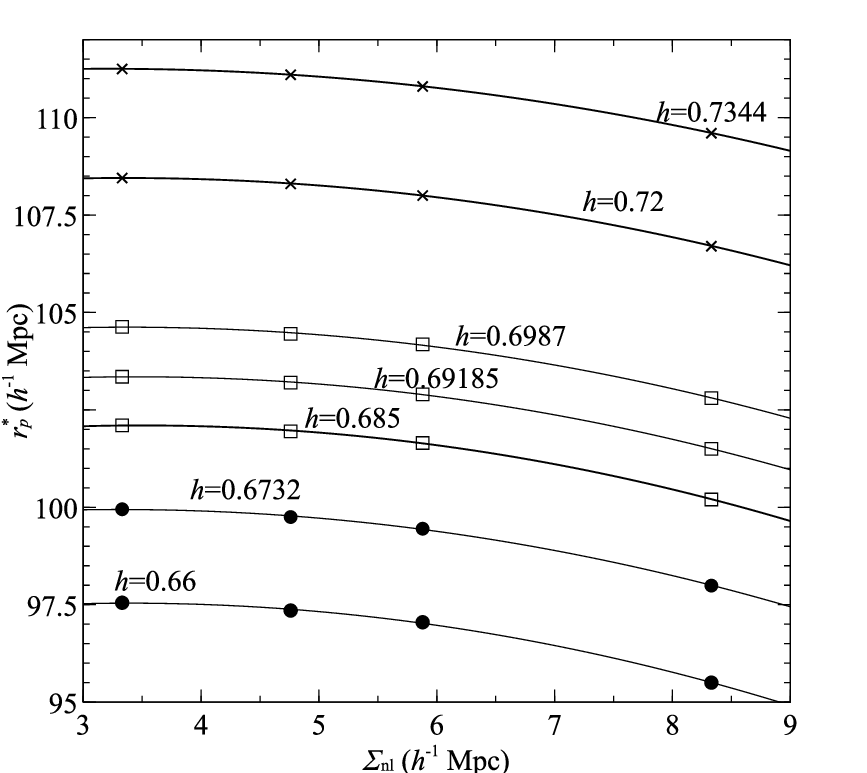}
\caption{ Location of the peak values of $\xi$$(r)$ plotted for the three test sets, a total of seven components, as determined from  Equation~\ref{E:corrfunc}, as a function of the value of the damping factor, $\Sigma_{nl}$ used in Equation~\ref{E:powspec}. The displayed points for all curves are at $\Sigma_{nl}$=3.33, 4.76. 5.88, and 8.33 $h^{-1}$ Mpc. Test set (1) is denoted by filled-in circles, test set (2) by open squares, and test set (3) by crosses.} 
\label{fig:Fig1}
\end{figure}

We introduce three multi-component sets, defined by different values of $h$, to establish calibration curves, which relate $h$ and the location of the BAO feature.  All sets, and subsequent analysis, make use of Equation~\ref{E:degen}, and the ancillary parameter values defined in the Introduction.  Since the $h$ value uniquely specifies the flat $\Lambda$CDM cosmology, we will generally refer to $h$ alone to identify a designated point. The three sets are (1) $h$=0.66 and 0.6732; (2) $h$=0.685, 0.69185, and 0.6987; and (3) $h$=0.72 and 0.7344.  In all three sets the highest component differs from the lowest component by 2 per cent.  Set (2) additionally has an element that differs from the lowest component by 1 per cent.   The selected values span the range from \textit{Planck} derived, to low-redshift derived results \citep{Aghanim2018,Riess2019}.  We use these sets to simulate BAO 2PCF solutions.

 The matter power spectrum, $P^{\text{lin}}(k)$, used in evaluating the correlation function, is generated with the CAMB software package   \footnote{Our CAMB calculations were derived from NASA's online Lambda utility at \url{https://lambda.gsfc.nasa.gov/toolbox/tb_camb_form.cfm}.} \citep{Lewis2000}. A baryon-free,``no-wiggle'', spectrum  \citep{Eisenstein1998a,Eisenstein1999}, $P^{\text{nw}}(k)$, is merged with  $P^{\text{lin}}(k)$ to comprise  a power spectrum, which suppresses higher order oscillations, thus regulating small scale non-linearities.  The modified matter power spectrum, $p^{\text{mod}}(k)$, is represented as, \citep{Percival2007},
\begin{equation}\label{E:powspec}
P^{\text{mod}}(k)=P^{\text{nw}}(k)+[P^{\text{lin}}(k)-P^{\text{nw}}(k)]e^{-(k\Sigma_{nl})^{2}/2}.
\end{equation}

The large co-moving radius of the BAO acoustic feature, $\sim$150 Mpc, leads to the feature being, for the most part, unaffected by the non-linearities initiated by the clustering of matter after recombination. However, these non-linearities, along with redshift distortions and galaxy biasing, do have some effect on the BAO correlation function peak location, $r_p$ \citep*{Guzik2007,Smith2008,Crocce2008}.  The absolute peak location shifts  by $\sim$0.5 to 1.0 percent as a result of these agencies.  We focus on the ratio of fiducial and data peak positions, 
\begin{equation}\label{E:alph} 
\alpha=r_{p,\text{fid}}/r_{p},
\end{equation}
 rather than on the absolute scale, and disregard the large-scale perturbations in our analysis. In doing so, our ratio of fiducial and data peak positions, $r^{*}_{p,\text{fid}}/r^{*}_{p}$  differs from $\alpha$ by no more than second order terms. As long as  $r^{*}_{p,\text{fid}}/r^{*}_{p}\simeq \alpha$, discounting the large-scale non-linear effects of the peak position is a valid premise.  The demonstration of this result is elementary, but because it is a key assumption, we establish its validity in Appendix~\ref{sec:appen1}.  That ability to subtract out the large-scale non-linearities without deleterious effect, contrasts with the need to retain the small-scale non-linearities through use of the damping parameter, $\Sigma_{nl}$ $h^{-1}$ Mpc, in describing the matter power spectrum.

  \begin{figure} 
\centering
 \includegraphics[width=\columnwidth]{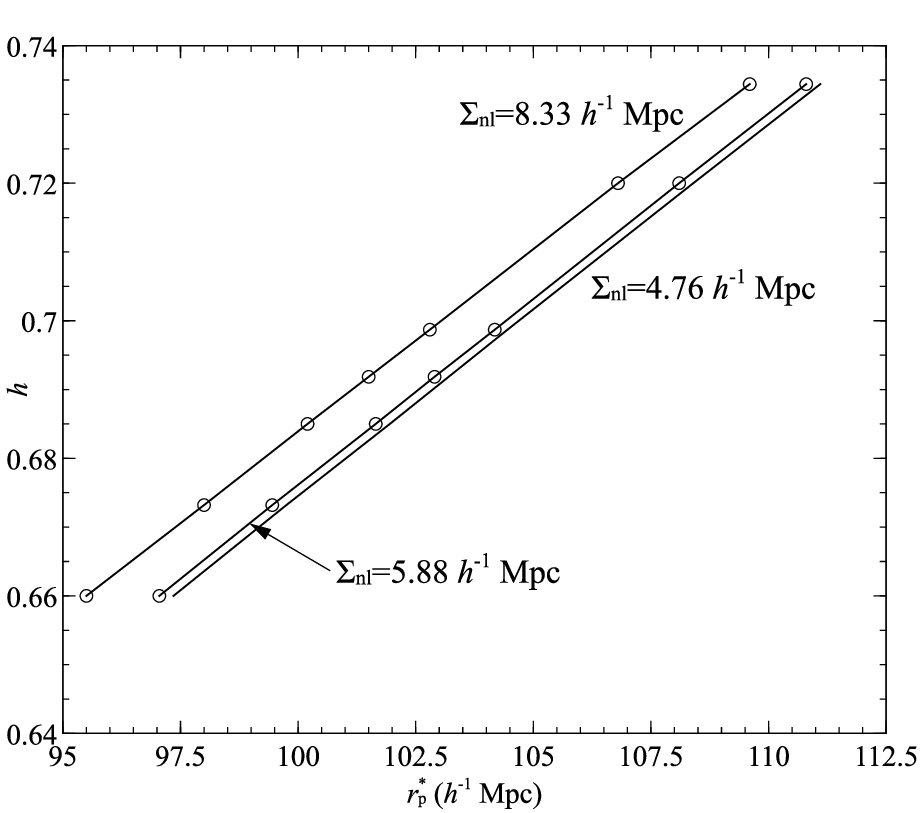}
\caption{Functional dependence of $h$ upon $r^{*}_{p}$ for three values of $\Sigma_{\text{nl}}$.  The fit line for $\Sigma_{nl}=4.76$ $h^{-1}$ Mpc, as an example, satisfies $h=0.005411r^{*}_{p} +0.1334$.  The open circles are the $h$ values for the components of the three test sets found in Figure~\ref{fig:Fig1}.  Open circles are not shown for $\Sigma_{\text{nl}}$=4.76 $h^{-1}$ Mpc, due to the close proximity of the lines. }
\label{fig:Fig2}
\end{figure}

 Computations have been performed at several values of $z$, dependent upon the  specific circumstance.  That is not a critical consideration. Though the 2PCF peak broadens with decreasing $z$, the peak position, $r_p^{*}$,  is independent of $z$. (See, e.g., \citet{Weinberg2013}, figure~11). The Fourier transform of Equation~(\ref{E:powspec}) \citep{Sanchez2008} produces the correlation function in real space,
\begin{equation}\label{E:corrfunc}
\xi(r)= \frac{1}{2 {\pi}^2}\int_{0}^{\infty} dk k^2 P^{\text{mod}}(k)j_{0}(kr)e^{-k^{2}a^{2}}.
\end{equation}
Here, $a^2$ is a damping factor that regulates convergence at high values of $k$ where the zero order spherical Bessel function, $j_{0}(kr)$ undergoes rapid oscillation.  Setting its value to 0.6 $h^{-2}~\text{Mpc}^2$, avoids interference with the determination of $r^{*}_p$.  The 2PCF peak location, $r^{*}_{p}$, is dependent upon the value of the $P^{\text{mod}}(k)$ damping parameter, $\Sigma_{nl}$. In Figure~\ref{fig:Fig1}, $r^{*}_{p}$ is plotted as a function of $\Sigma_{nl}$ for each component of the three test sets.  Those curves are valid over a range of values, but have been generated using $\Sigma_{nl}$=3.33, 4.76, 5.88, and 8.33 $h^{-1}$ Mpc, representing approximate extreme, and median values of $\Sigma_{nl}$. The value, 8.33 $h^{-1}$ Mpc, approaches the highest damping that can be introduced before the BAO feature disappears, morphing to a plateau. \citet{Padmanabhan2012} found a value of 8.1 $h^{-1}$ Mpc, pre-recombination. The low end, 3.33 $h^{-1}$ Mpc, is comparable to the transverse component of a composite damping factor used in Ly$\alpha$ studies \citet{Kirby2013}.  The value 4.76 $h^{-1}$ Mpc is the inverse median of those extremes, and 5.88 $h^{-1}$ Mpc is a computation, found in \citet{Beutler2011}, of a theoretical  determination \citep{Matsubara2008},  
\begin{equation}
\Sigma_{nl}=\bigg[\frac{1}{6\pi^{2}}\int_{0}^{\infty}dkP^{lin}(k)\bigg]^{1/2}.
\end{equation}
Figure~\ref{fig:Fig2} uses the same information as in Figure~\ref{fig:Fig1}, but in this instance we plot $h$ as a function of $r^{*}_{p}$. for three values of $\Sigma_{nl}$. (The fourth value, 3.33 $h^{-1}$ Mpc, is not plotted, for clarity. As seen from Figure~\ref{fig:Fig1}, the  dependence on $\Sigma_{nl}$ flattens at the low end, leading to dense packing of the fit lines.)  A linear fit is obtained for each of the values.  In the following section those fits, BAO study fiducial values, and $\alpha$ (or $\alpha_{\perp}$ for anisotropic investigations) are used to evaluate $h$ and $\Omega_{m}$ for several analyzed studies.  

\begin{figure} 
 \includegraphics[width=\columnwidth]{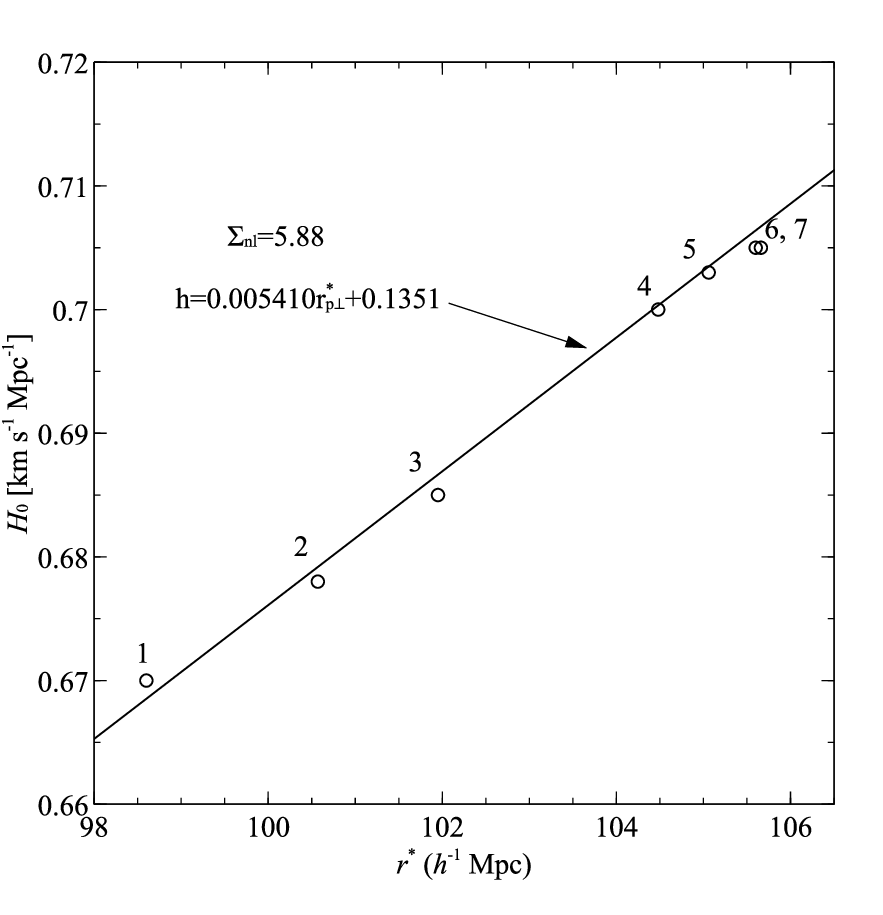}
\caption{Test of fit line for $\Sigma_{nl}=5.88 h^{-1}$ Mpc, $h=0.005410r^{*}_{p}+0.1351$. Values of $r^{*}_{p}$ have been computed using Table~\ref{tab:1} data point parameter inputs , and are numbered on the figure as per the table entries.} 
\label{fig:Fig3}
\end{figure}

Before proceeding, we first test this procedure with known quantities. In Figure~\ref {fig:Fig3}, the fit line for $\Sigma_{nl}=5.88 h^{-1}$Mpc is plotted along with data points from  three CMB studies, \citet{Calabrese2017,Ade2015}, and \textit{WMAP}9, \citet{Hinshaw2013}.  Table~\ref{tab:1} lists the parameter values for seven data points from those studies. Because \textit{WMAP} had limited resolution in comparison with \textit{Planck}, the ninth year, \textit{WMAP}9, measurements are supplemented in four of the determinations with data from the high resolution Atacama Cosmology Telescope (\textit{ACT}) and South Pole Telescope (\textit{SPT}) missions.  Two of those computations made use of both \textit{ACT} and \textit{SPT}, and are referred to as extended CMB, or eCMB.  The test set fit line is correlated with \textit{Planck} results through Equation~\ref{E:degen}.  \textit{Planck} data points extracted from the same data set \citep{Ade2015}, were analyzed by two different groups, and are included to illustrate the $\Omega_m$, $h$ degeneracy.  The other five data points are mostly independent of \textit{Planck} (\citet{Calabrese2017} incorporated the \citet{Ade2015} reionization optical depth, $\tau$, in their analysis).  The agreement in Figure~\ref {fig:Fig3} between the test set fit line and the data points is good enough that we emphasize that it is not a least square fit.

\begin{deluxetable*}{ccccccl} 
\tablenum{1}
\tablecaption{CMB parameter sets\label{tab:1}} 
\tablehead{\colhead{$\Omega_{b}h^2$} & \colhead{$\Omega_{c}h^2$} & \colhead{ $n_s$} & \colhead{$\tau$} & \colhead{$h$} & \colhead{$\Omega_m$} & \colhead{Ref.} }
\startdata
0.02217$\pm$0.00021&0.1205$\pm$0.0021&0.9625$\pm$0.0056& 0.064$\pm$0.01&0.670$\pm$0.009&0.319$\pm$0.013&(1) \citet{Calabrese2017}, \textit{Planck} \\
0.02226$\pm$0.00023&0.1186$\pm$0.0020& 0.9677$\pm$0.006&0.066$\pm$0.016&0.678$\pm$0.009&0.308$\pm$0.012&(2) \citet{Ade2015}\\
0.02243$\pm$0.00040 & 0.1156$\pm$0.0043&0.966$\pm$0.01&0.06$\pm$0.009&0.685$\pm$0.02&0.296$\pm$0.025&(3) \citet{Calabrese2017}, \textit{WMAP}9+\textit{ACT}\\
0.02264$\pm$0.00050 & 0.1138$\pm$0.0045&0.972$\pm$0.013&0.089$\pm$0.014&0.700$\pm$2.2&0.279$\pm$0.023&(4) \citet{Hinshaw2013}, WMAP9\\
0.02242$\pm$0.00032&0.1134$\pm$0.0036&0.9638$\pm$0.0087&0.058$\pm$0.009&0.703$\pm$0.016&0.276$\pm$0.019&(5) \citet{Calabrese2017}, \textit{WMAP}9+eCMB\\
0.02229$\pm$0.00037&0.1126$\pm$0.0035&0.9646$\pm$0.0098&0.084$\pm$0.013&0.705$\pm$1.6&0.272$\pm$0.017&(6) \citet{Hinshaw2013}, \textit{WMAP}9+eCMB\\
0.02223$\pm$0.00033&0.1126$\pm$0.0036& 0.9610$\pm$0.0089& 0.057$\pm$0.009&0.705$\pm$0.016&0.273$\pm$0.019&(7) \citet{Calabrese2017}, \textit{WMAP}9+\textit{SPT}\\
\enddata
\tablecomments{Neutrino mass, $\Sigma  m_{\nu}=0.06$ ev for data sets 1, 2, 3, 5, and 7.  For data sets 4, and 6, the value is zero.  $\Omega_{c}$ is the cold dark matter fraction.  \citet{Calabrese2017}, \textit{Planck}, reference (1) refit  the data from \citet{Ade2015}, Reference (2). The difference in the $\Omega_m$, $h$ pair values is indicative of the $\Omega_m$, $h$ pair  degeneracy in CMB TT spectra parameter evaluations.}
\end{deluxetable*}

 \section{Analysis}
 \label{sec:anal}
 
 \begin{figure*} 
 \centering
 \includegraphics[scale=.65]{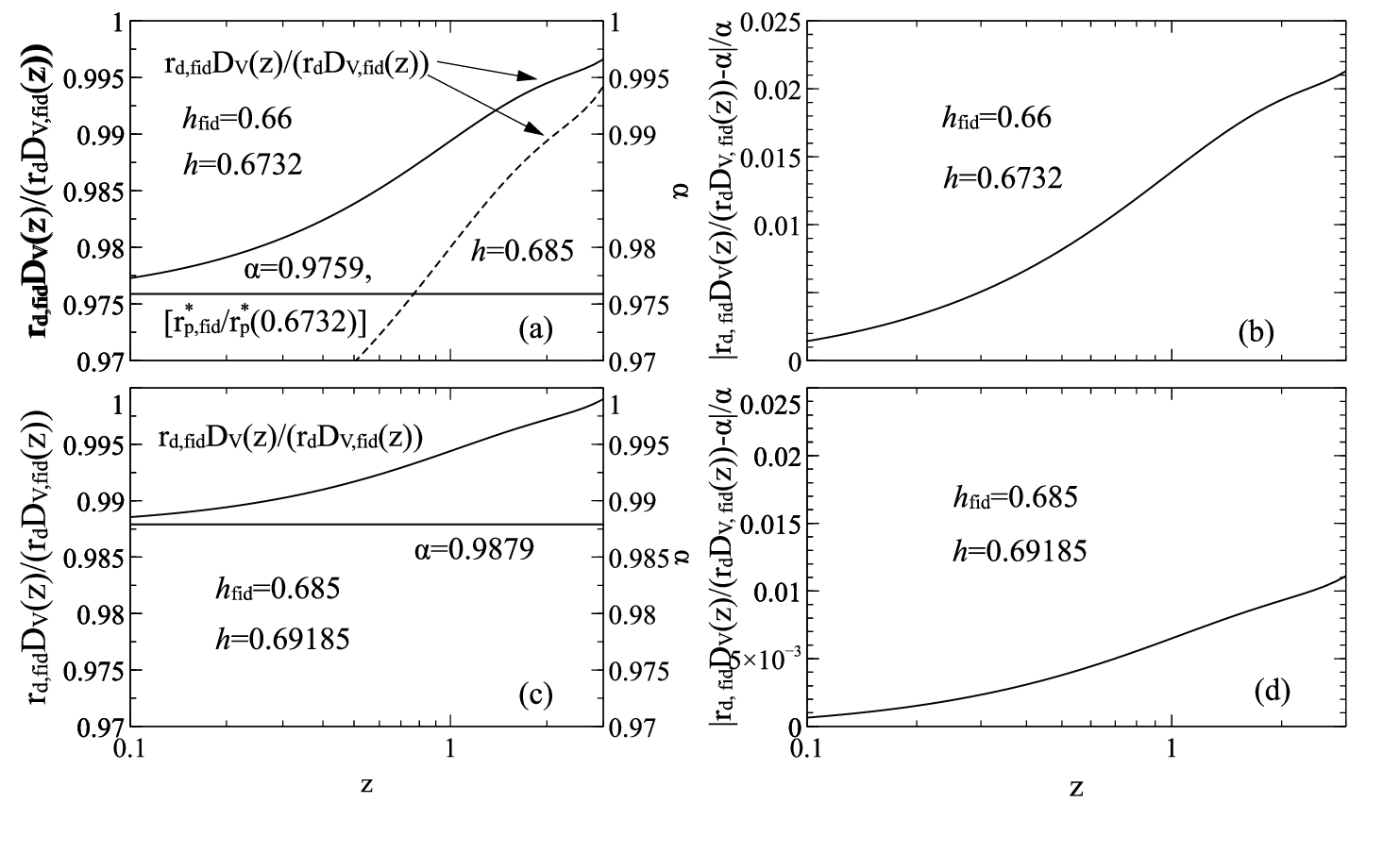}
\caption{ Both sides of Equation~\ref{E:DV3} are plotted in (a) and (c) to assess the deviation from equality. Values of $r^{*}_{p}$, used to evaluate $\alpha$, are obtained from Figure~\ref{fig:Fig1}, with $\Sigma_{nl}$=4.76. The normalized deviations are shown in the righthand boxes.  An example of the error generated by the deviations is given in box (a), where a value of $h$=0.685 would be obtained using Equation~\ref{E:DV3} for a measurement at $z\simeq$ 0.75, rather than the correct value of $h$=0.6732. To avoid ambiguity, the value of $\alpha=0.9759$ in (a) is specifically indicated as denoting the ratio of $r^{*}_{p, \text{fid}}/r^{*}_{p}$ at $h$=0.6732} 
\label{fig:Fig4}
\end{figure*}

  \citet{Eisenstein2005} instituted the volume averaged distance construct, 
\begin{equation}\label{E:DV2}
D_V(z)= \left( D_M^{2}(z) \frac{cz}{H(z)} \right)^{1/3},
\end{equation}
where $D_{M}(z)$ is the comoving angular diameter distance, and $H(z)$ is the Hubble parameter.  The function $D_V$(\textit{z}) is designed to compensate for the distortion of radial distance in redshift space.  For those analyses in which the correlation function, $\xi (r)$, is spherically averaged, and an isotropic distance scale is extracted, the BAO feature is characterized  by $\alpha$, (Eq.~{\ref{E:alph}}). 

 \citet{Anderson2012,Reid2012,Sanchez2012} introduced the equation,
 \begin{equation}\label{E:DV1}
\frac{D_{V}(z)}{r_{d}}=\alpha \frac{D_{V,\text{fid}}(z)}{r_{d,\text{fid}}}, 
\end{equation}
where $r_d$ is the comoving acoustic horizon at the drag epoch.  That form has since been universally used for isotropic BAO studies \citep{Padmanabhan2012,Mehta2012,Anderson2014,Kazin2014,Cuesta2016}.   The analogous equations used in anisotropic BAO studies are,
\begin{equation}  
\label{E:Ma}
 \frac{D_{M}(z)}{r_{d}}=\alpha_{\perp} \frac{D_{M,\text{fid}}(z)}{r_{d,\text{fid}}}, 
 \end{equation}
 \begin{equation}
  \label{E:Ha}
 \frac{D_{H}(z)}{r_{d}}=\alpha_{\|} \frac{D_{H,\text{fid}}(z)}{r_{d,\text{fid}}},
 \end{equation}
where $D_{H}(z)=c/H(z)$.  

The analysis for the anisotropic correlation function typically splits the BAO feature into two components, $r_{p\perp}$ and $r_{p\parallel}$, the transverse and radial components, respectively.   As with the isotropic relationship, $\alpha=r_{p,\text{fid}}/r_{p}$, the relationships, $\alpha_{\perp}=r_{p\perp ,\text{fid}}/r_{p\perp}$, and $\alpha_{\parallel}=r_{p\parallel ,\text{fid}}/r_{p\parallel}$ describe the anisotropic results.  Because of redshift distortion, the measured values of, $\alpha$, $\alpha_{\parallel}$ and $\alpha_{\perp}$ differ from each other.  The relationships between the various dilation/contraction factors are \citep{Cuesta2016},
\begin{equation}
\label{E:parper}
\alpha=\alpha_{\parallel}^{1/3}\alpha_{\perp}^{2/3} 
\end{equation}
\begin{equation}
1+\epsilon=\bigg(\frac{\alpha_{\parallel}}{\alpha_{\perp}}\bigg)^{1/3},
\end{equation}
where $\epsilon$ is a distortion parameter.  With $\epsilon$=0, $\alpha=\alpha_{\parallel}=\alpha_{\perp}$.  In calculating the  fiducial value, $r_{p, \text{fid}}^{*}$, redshift distortion is not a consideration, $r_{p, \text{fid}}^{*}\equiv r_{p\perp, \text{fid}}^{*}$. Similarly, the results of Figures~\ref{fig:Fig1} and \ref{fig:Fig2} carry over to the anisotropic analysis.

We evaluate $r_d$ using an expression  devised by \citet{Aubourg2015} that approaches CAMB derived values to within 0.021 per cent, when values of $\omega_{cb}$ and $\omega_b$  are within 3$\sigma$ of the \textit{Planck} derived values: 
\begin{equation}
\label{E:rd}
r_d\simeq\frac{55.154~\text{exp}[-72.3(\omega_{\nu}+0.0006)^2]}{\omega_{cb}^{0.25351}\omega_{b}^{0.12807}} ~\text{Mpc},
\end{equation}
where $\omega_{x}\equiv\Omega_{x}h^2$.  The subscript, $cb$, pertains to the total mass fraction of CDM plus baryons.  We assume $\Omega_{\nu}$$h^2$=0.00064.

 Equations~{\ref{E:DV1}}, {\ref{E:Ma}}, and {\ref{E:Ha}} were not derived from first principles.  They are algorithms that require verification. In Appendix \ref{sec:appen} we demonstrate that Equations~\ref{E:DV1}, and \ref{E:Ma} provide excellent descriptions of the cosmology as $z \Rightarrow 0$.  Because of redshift space distortion, we do not consider Equation~\ref{E:Ha}.  Equation~\ref{E:DV1} also exhibits red shift distortion.  However, since the early BAO studies were all isotropic, we explore those studies in Section~\ref{sec:iso}, and then take a second look in Subsection~\ref{sec:iso2}.  
 
 At the other extreme, at the drag redshift, $z_{d}\simeq$ 1060.  \begin{equation}  
\label{E:zdrag}
 \frac{D_{M, \text{fid}}(1060)}{r_{d, \text{fid}}}\simeq \frac{D_{M}(1060)}{r_{d}}. 
\end{equation}
Thus, the dependence upon $\alpha_{\perp}$ in Equation~{\ref{E:Ma}} is not sustainable at higher values of  $z$.  As an illustration of the validity of Equation~\ref{E:zdrag}, consider the situation at the drag redshift for two values of $h$.  For $h$=0.6732 $z_d$ =1059.82, $D_{M}(1059.82)/r_{d}$= 94.72, while for $h$=0.69185, $z_d$ =1059.55, $D_{M}(1059.55)/r_{d}$=94.59, a difference of $\sim$0.1 per cent. Values for $D_M$ are obtained from the Cosmotools online calculator.\footnote{The Cosmotools calculator is found online at, \url{http://www.bo.astro.it/~cappi/cosmotools}. }   The effect of fixing $z$=1060 for both values of $h$ is inconsequential.  Simulating a BAO algorithm solution, we take $h_{\text{fid}}$=0.6732, and $h$=0.69185, with $\Sigma_{nl}$=4.76, and find $\alpha_{\perp}=0.9666$, and $\alpha_{\perp}D_{M,\text{fid}}(1060)/r_{d,\text{fid}}=91.55$.  That differs from $D_{M}(1060)/r_{d}$ by over 3 per cent.  Our computations, in what follows, further bear out this transition from the BAO algorithms to Equation~\ref{E:zdrag} with increasing $z$.

 \subsection{Isotropic}
 \label{sec:iso}
  
 We rewrite Equation~{\ref{E:DV1}} as,

\begin{equation}  
\label{E:DV3}
\alpha=\frac{r_{d, \text{fid}}D_{V}(z)}{r_{d}D_{V, \text{fid}}}.
\end{equation}
Here the left side of the equation is the measured value of $\alpha$, while the right side is the kernel of the algorithm that is being tested. We plot both sides of the equation in Figure~\ref{fig:Fig4}a as a function of $z$.  This simulation uses the two components of test set (1), $h$=0.66, and 0.6732, with 0.66 arbitrarily  chosen as the fiducial value.  
$D_{V}(z)$ and $D_{V,\text{fid}}(z)$ are evaluated using Equation~\ref{E:DV2}. As throughout this manuscript, calculations are performed for a flat cosmology, with $w=-1$, and therefore,
\begin{equation}\label{E:Hz}
H(z)=H_{0}[\Omega_{m}(1+z)^3+1-\Omega_{m}]^{1/2}~\text{km}~ \text{s}^{-1} \text{Mpc}^{-1}.
\end{equation}

Figure~\ref{fig:Fig4}a shows distinct deviation between the two sides of Equation~\ref{E:DV3} with increasing redshift.  The difference between $\alpha$ and $r_{d, \text{fid}}D_{V}(z)/(r_{d}D_{V, \text{fid}}(z))$ normalized to $\alpha$ is depicted in Figure~\ref{fig:Fig4}b, and is about 1 per cent at $z\simeq$ 0.75.  Since $\alpha$ would be the measured quantity, and the fiducial values are a given, the deduced value of $h$ from Figure~\ref{fig:Fig4}a at $z\simeq$ 0.75, as indicated on the figure, and when using  Equation~\ref{E:DV3}, is $h\simeq0.685$, rather than the actual value of $h=0.6732$.  We find similar results for the $h$ pairs for the other two test sets, where $h_{\text{fid}}$ and $h$ differ by two per cent. In Figure~\ref{fig:Fig4}c and d we plot the pair from test set (2) in which $h_{\text{fid}}$ and $h$ differ by 1 per cent, and, as to be expected, find a substantial reduction in the deviation, a one per cent deviation, being found here at $z\simeq 3$.

In Figures~\ref{fig:Fig4}a and \ref{fig:Fig4}c note that the right hand side of Equation~\ref{E:DV3} has made significant progress towards unity at the coordinate maximum of $z$=3, far from the drag value of $\sim 1060$. That follows since, at $z\sim 3$, for the flat $\Lambda$CDM cosmologies considered here, the age of the Universe was $\sim$ 2.2 Gyr, and thus the transition of the right side of Equation~\ref{E:DV3} from near equality with $\alpha$ to near equality with unity, as dictated by Equation~\ref{E:zdrag}, occurs at a much lower redshift value than might at first be anticipated.

BAO 2PCF results from 2005 to date can be characterized as first generation, in that the standard deviations are typically several per cent. The \citet{Alam2017} combined measurements, presented in several papers based on the SDSS-III Baryon Oscillation Spectroscopic Survey (BOSS), represent the highest attained first generation precision, of $\sim$ 1 per cent. For surveys conducted at low values of redshift, such as \citet{Beutler2011} at $z=0.106$, and \citet{Ross2015} at $z=0.15$ the Equation~\ref{E:DV1} algorithm is a good fit.  Beyond those two low redshift figures, BAO galactic studies span a range of 0.32 $\eqslantless$ $z$ $\eqslantless$ 0.73.  We will not systematically appraise those studies. However, we do examine two isotropic studies, \citet{Cuesta2016}, because of its 1 per cent precision, and \citet{Ata2018}, because of its high redshift.  \citet{Cuesta2016}, like \citet{Alam2017}, was based on the Data Release 12 (DR12) of the SDSS BOSS program, and summary figure 13 of \citet{Alam2017} indicates that it is a good representation of those studies.  Measurements were conducted at effective redshifts $z$=0.32, and 0.57. We address the $z$=0.57 results. \citet{Ata2018} evaluated measurements on a population of quasars centered at a redshift of 1.52.  

\begin{figure} 
\centering
\includegraphics[width=\columnwidth]{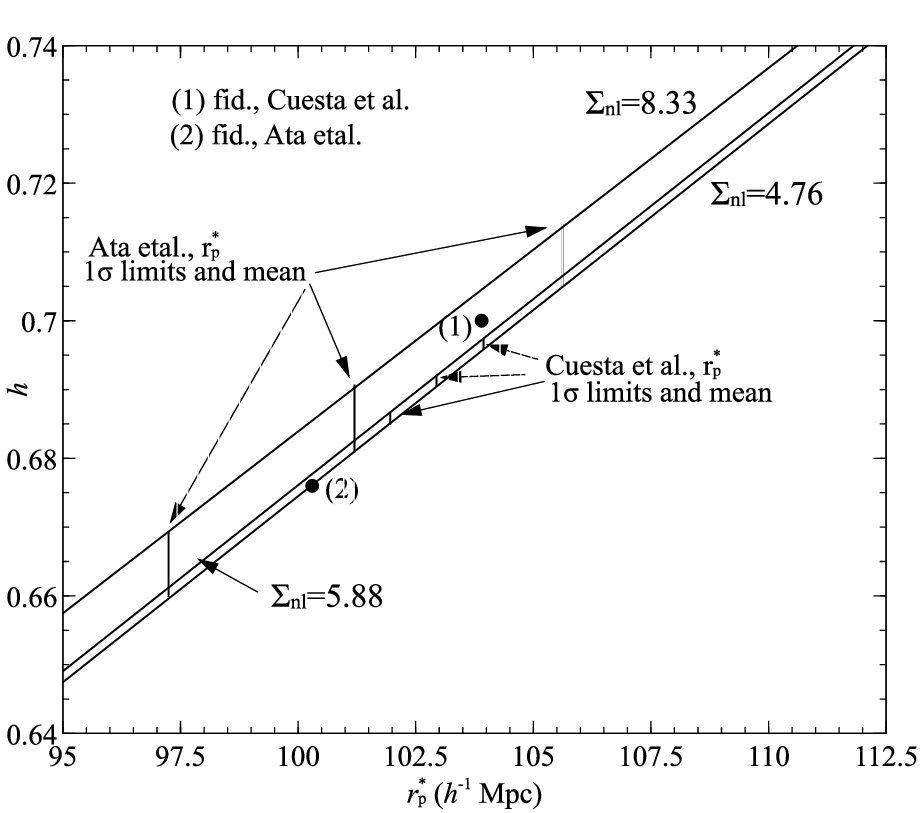}
\caption{ The fit lines of Figure~\ref{fig:Fig2} are reproduced here, and the loci of the \citet{Cuesta2016}, and \citet{Ata2018} $h$ limits indicated, as determined from Equations~\ref{E:h476} and \ref{E:h588}. The vertical lines delimit the extent of the damping parameter uncertainty.} 
\label{fig:Fig5}
\end{figure}

In Figure~\ref{fig:Fig5} we reproduce the fit lines of Figure~\ref{fig:Fig2}. That is not strictly correct for  application to specific BAO studies, since they include the large-scale perturbations that we have neglected.  Including those perturbations would slightly shift the fit lines, however, as in Appendix~\ref{sec:appen1}, the difference between the two results is of second order.   By expressing $r^{*}_{p}$ as $r^{*}_{p, \text{fid}}/\alpha$, the value of $h$ is derived with known quantities. Thus, for  $\Sigma_{\text{nl}}=4.76$ $h^{-1}$ Mpc, 
\begin{equation}
\label{E:h476}
h=0.005411r^{*}_{p,\text{fid}}\big/\alpha+0.1334, 
\end{equation}
and for $ \Sigma_{\text{nl}}=5.88$ $h^{-1}$ Mpc,
\begin{equation}
\label{E:h588}
h=0.005410r^{*}_{p,\text{fid}}\big/\alpha +0.1351.
\end{equation}
The fit line equation for $\Sigma_{\text{nl}}=8.33$ $h^{-1}$ Mpc is,
 \begin{equation}
 \label{E:h833}
h=0.005285r^{*}_{p,\text{fid}}\big/\alpha+0.1554. 
\end{equation}

\begin{deluxetable}{cccccl} 
\tablenum{2}
\tablecaption{Fiducial parameter sets for BAO studies discussed in this paper\label{tab:2}} 
\tablehead{\colhead{$\Omega_{b}h^2$} & \colhead{$n_s$} & \colhead{$\sigma_{8}$\tablenotemark{a}}& \colhead{$h$} & \colhead{$\Omega_m$} & \colhead{Ref.}} 
\startdata
0.02247&0.97& 0.8&0.7&0.29&\citet{Cuesta2016}\\
0.022&0.97&0.8&0.676&0.31&\citet{Ata2018}\\
0.02222 & 0.9655&0.830&0.6731&0.3147&\citet{Bautista2017} \\
0.0227&0.97&0.8&0.7&0.27& \citet{Delubac2015}\\
\enddata
\tablenotetext{a}{We have performed CAMB computations with $\tau $=0.054, rather than with $\sigma_8$. The value of $\sigma_8$ squared is proportional to the scalar amplitude. Maintaining all other fiducial values, changing the scalar amplitude does not affect the position of the BAO feature.}
\tablecomments{The neutrino mass, $\Sigma  m_{\nu}=0.06$ ev for all data sets, with the exception of \citet{Cuesta2016}, where it is set equal to zero. \citet{Bourboux2017,deSainte2019,Blomqvist2019} use the same fiducial values as \citet{Bautista2017}, and \citet{Font2014} use the same values as \citet{Delubac2015}.}
\end{deluxetable}

 All fiducial values for studies discussed herein can be found in Table~\ref{tab:2}.  For \citet{Cuesta2016} $\Sigma_{\text{nl}}=5.0$ $h^{-1}$ Mpc (see \citet{Vargas2018}, for which our value of 4.76 is adequate.  We depict $r^{*}_{p,\text{fid}}$=103.9 $h^{-1}$ Mpc on the plot. Note that it does not fall on the fit line, since the fiducial parameters do not satisfy the $h$, $\Omega_m$ near degeneracy condition of Equation~{\ref{E:degen}}.   We assume that the derived data do satisfy the near degeneracy condition. The study measured $\alpha$=1.0093$\pm$0.0097, which from Equation~{\ref{E:h476}} leads to $h$=0.6904$\pm$0.0058. While the fiducial value of  $\Sigma_{nl, \text{fid}}$ is a given, we do not know the actual damping parameter value. \citet{Cuesta2016} performed their analysis before and after implementing the reconstruction technique initiated by \citet{Eisenstein2007}.  That method employs the galaxy density data to compute the velocity field and then, effectively, runs the clock backwards,  reversing the inflowing mass. We use the \citet{Cuesta2016} post-reconstruction results, in which case we can dismiss the likelihood of higher damping levels \citep{Padmanabhan2012}.  Thus the narrow damping parameter uncertainty indicated in the plot.
  
 \citet{Ata2018} assign $\Sigma_{\text{nl}}=6.0$ $h^{-1}$ Mpc,  which is adequately approximated by our $\Sigma_{\text{nl}}=5.88$ $h^{-1}$ Mpc.  They found $\alpha$=0.991$\pm$0.037. Their fiducial parameters return $r^{*}_{p,\text{fid}}$=100.4 $h^{-1}$ Mpc, which, from Equation~\ref{E:h588}, leads to $h$=0.6832$\pm$0.022. Because of the low density of quasars, \citet{Ata2018} were not able to perform reconstruction.  Therefore in Figure~\ref{fig:Fig5} we depict a large damping parameter uncertainty, extending from $\Sigma_{\text{nl}}=8.33$ $h^{-1}$ Mpc down to the region where the fit lines coalesce. At the mean value of $\alpha$, with regard to Equation~\ref{E:DV3}, the normalized deviation is $\mid$$\alpha$ - $r_{d, \text{fid}}D_{V}(z)/(r_{d}D_{V, \text{fid}}(z))$$\mid$$\big/$$\alpha$=0.008.   Considering the $\alpha$ standard deviation of $\sim$3.7 per cent, the bias introduced by using Equation~\ref{E:DV3} is not appreciable.

Though \citet{Cuesta2016} achieved a precision where deviations could be of import, the relatively low redshift of $z$=0.57, and a value of $\alpha$ close to unity, minimized the negative influence. Because of the predominantly large variances of first generation isotropic BAO studies, the deviations from equality in Equation~\ref{E:DV3} do not, in general, adversely affect the results.  The proviso to that conclusion is that the deviations are not errors to be added in quadrature, but rather, with respect to Equation~\ref{E:DV3}, are unwanted biases that are directly, and misleadingly, added to  flat $\Lambda$CDM consistent results. We return to the subject of isotropic BAO studies in Subsection~\ref{sec:iso2}. 

\subsection{Anisotropic}
 \label{sec:anis}
 
  \begin{figure} 
\centering
 \includegraphics[width=\columnwidth]{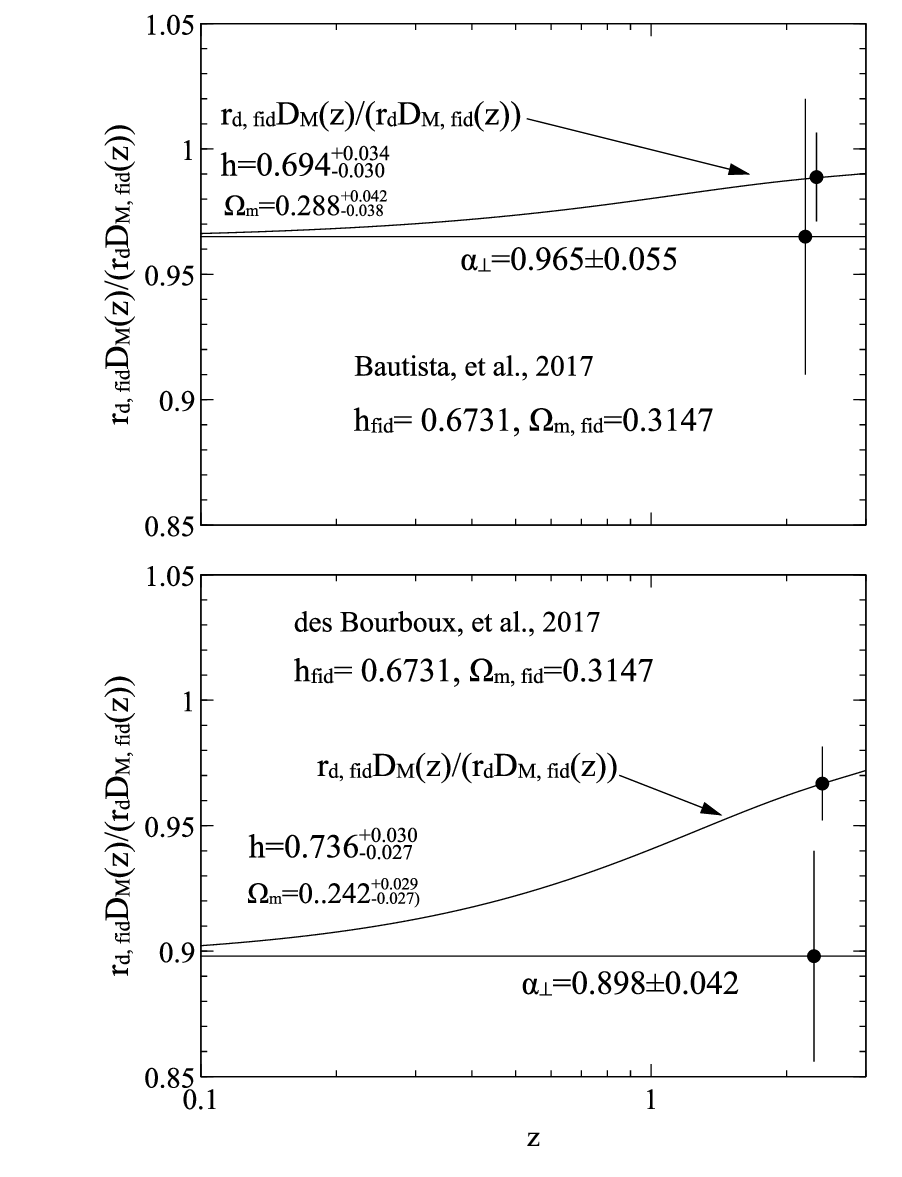}
\caption{ Both sides of Equation~\ref{E:DM1} are plotted in the upper and lower panels for  \citet{Bautista2017} at $z$=2.33, and for \citet{Bourboux2017} at $z$=2.4, respectively, to assess the deviation from equality. The standard deviation bars for $\alpha_{\perp}$ are shifted to the left for clarity.  } 
\label{fig:Fig6}
\end{figure}
 
Anisotropic determinations \citep{Xu2013,Kazin2013,Anderson2014,Anderson2014b,Cuesta2016} have been facilitated as spectroscopic galaxy surveys have increased in size. They provide more detail than the isotropic average, and when taken together with reconstruction \citep{Xu2013} contribute to a more nuanced understanding of BAO conditions.
Although reconstruction significantly ameliorates the effects of redshift distortion \citep{Weinberg2013}, those effects are not completely eliminated.  The most direct measure of the BAO feature is provided by $\alpha_{\perp}$.  Redshift distortion is not a factor here, and $r_{p\perp}$ is a direct measure in real space as well as in redshift space. For that reason, in what follows, we concentrate on that transverse component.

\subsubsection{Ly$\alpha$ forest anisotropic studies}
\label{sec:ly}

\begin{figure} 
\centering
 \includegraphics[width=\columnwidth]{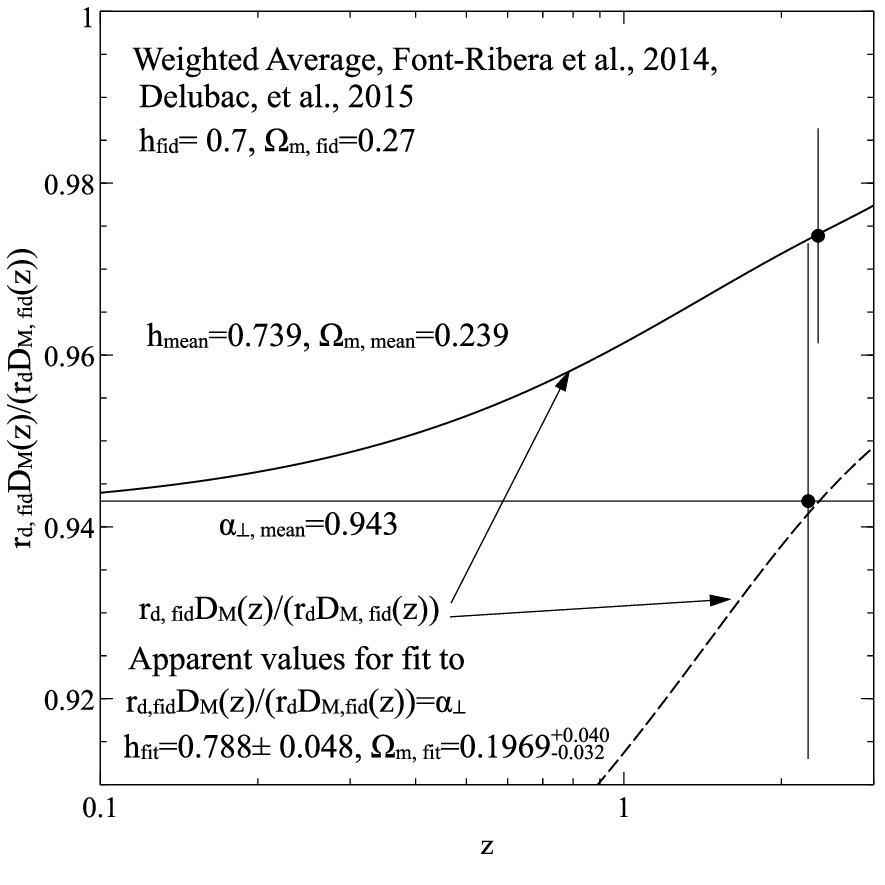}
\caption{ Both sides of Equation~\ref{E:DM1} are plotted for the weighted average $\alpha_{\perp}=0.943\pm0.030$ values of the  \citet{Delubac2015,Font2014} Ly$\alpha$ forest studies. The fiducial values listed in Table~\ref{tab:2} yield $r_{p, \text{fid}}^{*}$=105.7, and the flat $\Lambda$CDM solution is $h=0.739\pm0.020$. The mean normalized discrepancy is $\mid$$\alpha_{\perp}$-$r_{d, \text{fid}}D_{M}(z)/(r_{d}D_{M, \text{fid}}(z))$$\mid$/$\alpha_{\perp}$$\simeq$0.033. Apparent values to satisfy Equation~\ref{E:DM1} are $h=0.788\pm0.048$, and $\Omega_{m}=0.1969^{+0.040}_{-0.032}$. The standard deviation bars for $\alpha_{\perp}$ are shifted to the left for clarity. } 
\label{fig:Fig7}
\end{figure}

\begin{deluxetable*}{llccll}
\tablenum{3}
\tablecaption{Comparison of anisotropic and isotropic computations of $\Omega_{m}$ and $h$\label{tab:anis}} 
\tablehead{\colhead{Ref.} & & \colhead{ $\alpha_{\perp}$} & \colhead{$\Omega_{m}(\alpha_{\perp})$ } & \colhead{ $\alpha$ } & \colhead{$\Omega_{m}(\alpha)$} \\
\colhead{Survey} & \colhead{Effec. $z$} & \colhead{$\alpha_{\parallel}$} & \colhead{$h(\alpha_{\perp})$} & \colhead{$\rho_{\alpha_{\perp} \alpha_{ \parallel}}$} & \colhead{$h(\alpha$)}}
\startdata
\citet{Cuesta2016} & 0.57 & 1.0368$\pm$0.0142 &0.312$\pm$0.011 & 1.0051$\pm$0.0098 &0.290$\pm$0.070\\
BOSS DR12&&0.9446$\pm$0.0324&0.676$\pm$0.008&-0.5671&0.693$\pm$0.006\vspace{0.7pt}\\
\citet{Blomqvist2019}&2.35&0.923$\pm$0.046&$0.258^{+0.034}_{-0.031}$& 0.971$\pm$0.028& 0.293$\pm$0.022\vspace{1.5pt}\\
eBOSS DR14&&1.076$\pm$0.042&$0.720^{+0.032}_{-0.030}$&-0.44&$0.690^{+0.018}_{-0.017}$\vspace{1.5pt}\\
\citet{deSainte2019}&2.34&0.953$\pm$0.048&$0.280^{+0.038}_{-0.034}$& 0.979$\pm$0.03&0.299$\pm$0.024\vspace{1.5pt}\\
eBOSS DR14 &&1.033$\pm$0.034&$0.701^{+0.032}_{-0.029}$&-0.34&$0.686^{+0.019}_{-0.018}$\\	
\enddata
\tablecomments{Isotropic ratio, $\alpha$, is computed using Equation~\ref{E:parper}. Fiducial parameters for \citet{Cuesta2016} are the same as for the isotropic analysis. Fiducial values for \citet{deSainte2019} and \citet{Blomqvist2019} are identical to those of \citet{Bautista2017} and  \citet{Bourboux2017}. For \citet{Cuesta2016}, the computed value from Equation~\ref{E:parper} compares favorably with their isotropic measurement of $\alpha$=1.0093$\pm$0.0097.}
\end{deluxetable*}

Ly$\alpha$ forest BAO measurements are undertaken at redshifts, $z>2$, and are therefore  of specific interest in this study. The auto-correlation Ly$\alpha$ forest analysis of \citet{Bautista2017} and the cross-correlation study of \citet{Bourboux2017} cover the same segments of sky, but their  $\alpha_{\perp}$, $\alpha_{\parallel}$, pairs are largely uncorrelated \citep{Bourboux2017}.  

The Ly$\alpha$ forest computations model the damping parameter as \citep{Kirby2013},
\begin{equation}
\Sigma^{2}(\mu_{k})= \mu_{k}^{2}\Sigma_{\parallel}^{2}+(1-\mu_{k}^{2})\Sigma_{\perp}^{2}.
\end{equation}
Here, $\mu_{k}$ is the cosine of the angle between the radial direction and the line joining two tracers `of the large-scale distribution of matter in redshift space'. $\Sigma_{\parallel}=6.41$ $h^{-1}$ Mpc, and $\Sigma_{\perp}=3.26$ $h^{-1}$ Mpc. We use our $\Sigma_{nl}$ value of 3.33 $h^{-1}$ Mpc, which is close to the $\Sigma_{\perp}$ value of 3.26 $h^{-1}$ Mpc. The fit line for $\Sigma_{\text{nl}}=3.33$ $h^{-1}$ Mpc  is,
\begin{equation}
\label{E:h333}
h=0.005431r^{*}_{p\perp,\text{fid}}\big/\alpha_{\perp}+0.1304
\end{equation}
From the fiducial values we find, $r^{*}_{p, \text{fid}}$=100.13 $h^{-1}$ Mpc.  Analogous to Equation~\ref{E:DV3}, we rewrite Equation~\ref{E:Ma} as,
\begin{equation}  
\label{E:DM1}
\alpha_{\perp}=\frac{r_{d, \text{fid}}D_{M}(z)}{r_{d}D_{M, \text{fid}}},
\end{equation}
and plot both sides of the equation in Figure~\ref{fig:Fig6}. Although each study is represented by one point on the plot,  \citet{Bautista2017}, at $z$=2.33 in the top panel, and  \citet{Bourboux2017}, at $z$=2.4 in the lower panel, we have plotted the mean values of $\alpha_{\perp}$ and $r_{d, \text{fid}}D_{M}(z)/(r_{d}D_{M, \text{fid}}(z))$ over a range of redshifts, $0.1<z<3.0$ to give an indication of the deviation from equality as $z$ increases.  For  \citet{Bautista2017} we find $h=0.694^{+0.036}_{-0.032}$, and for \citet{Bourboux2017}, $h=0.736^{+0.031}_{-0.029}$.  These values differ significantly from the fiducial values, selected by the authors to coincide with \citet{Ade2015}.  The deviations of Equation~\ref{E:DM1} from equality would move the apparent values even further from the \textit{Planck} likelihood.  

The \citet{Bautista2017} and \citet{Bourboux2017} measurements have been updated by \citet{deSainte2019} and \citet{Blomqvist2019} using the same fiducial parameters.   \citet{Blomqvist2019} provides a combined fit value for both studies of $\alpha_{\perp}=0.942^{+0.032}_{-0.030}$, from which we find $h=0.708\pm0.02$.

\citet{Aubourg2015,Addison2018,Aghanim2018} have all noted the discrepancy between the Ly$\alpha$ forest BAO measurements and the \textit{Planck} evaluation, with the former consistently finding higher values of $h$.  We have also found that to be true, but additionally the \citet{Aubourg2015,Addison2018,Aghanim2018} investigations used the BAO Equations~\ref{E:Ma}, and \ref{E:Ha}.  Therefore, we expect that  they would find an even larger disparity.  In addition to performing joint fits between BAO and \textit{Planck} data, \citet{Addison2018} also considered constraints from  the BAO scale alone.  For that undertaking they combined earlier Ly$\alpha$ forest studies, \citet{Delubac2015,Font2014}.  In Figure~\ref{fig:Fig7} we plot our analysis of the weighted average of those two studies. The data point is at $z$=2.35.  We find $h=0.739\pm0.020$. The apparent value of $h$ that satisfies Equation~\ref{E:DM1} is $0.788\pm0.048$, with $\Omega_{m}=0.1969^{+0.040}_{-0.030}$. By employing Equation~\ref{E:DM1}, \citet{Addison2018}, figure 3, elicited a similar result, providing a clear-cut example of the misleading nature of Equations~\ref{E:Ma} and \ref{E:DM1} at higher values of $z$.

\subsubsection{Assessment of isotropic measurements}
\label{sec:iso2}

Separate from our analysis of the applicability of the BAO equations, we turn back to consideration of isotropic results in light of what is observed in anisotropic measurements.  Table~\ref{tab:anis} lists $\alpha_{\perp}$, $\alpha$, and respective derived values of $\Omega_{m}$ and $h$ for the most recent Ly$\alpha$ forest studies, \citet{deSainte2019} and \citet{Blomqvist2019}, plus the anisotropic results of \citet{Cuesta2016}, earlier considered regarding isotropic outcomes. The isotropic ratio, $\alpha$, is computed using Equation~\ref{E:parper}.  As previously noted, $\alpha_{\perp}$ provides the most reliable measure, in that it is free of the effects of redshift distortion.  For each of the three studies, there is a pronounced difference between the $h$ and $\Omega_{m}$ results as determined by $\alpha_{\perp}$, versus those determined by $\alpha$. Because of this significant dissimilarity, we question the viability of using isotropic study results for cosmological parameter assessments.  Specifically, \citet{Beutler2011} and \citet{Ross2015} have both been used in the \citet{Aubourg2015,Addison2018,Aghanim2018} investigations.  Because of their low redshifts, we concluded that there was no problem with using Equation~\ref{E:DV1} to describe these results.  However, with increased precision, and the ability to perform anisotropic measurements, use of these earlier isotropic results is unwarranted.

  \section{Conclusions}
 \label{sec:Conclusions}
 
The BAO Equations~\ref{E:DV1} and \ref{E:Ma} are algorithms, which, as demonstrated in Appendix~\ref{sec:appen}, are accurate when $z\Rightarrow 0$, and, as established in the  text, are adequate at low values of $z$, in particular for the early BAO studies, with high variances. (Because of redshift distortion, the third algorithm of the set, Equation~\ref{E:Ha} was not considered.) We have focused on the transition of the quantity,  $r_{d, \text{fid}}D_{M}(z)/(r_{d}D_{M, \text{fid}}(z))$ from $\alpha_{\perp}$ as $z\Rightarrow 0$, to $\sim$ unity as $z\Rightarrow z_{d} \simeq 1060$.  At the lower limit, the transverse anisotropic BAO algorithm provides an excellent fit.  However, with increasing $z$ that equation deviates from an accurate description. Figures~\ref{fig:Fig4}, \ref{fig:Fig6} and \ref{fig:Fig7} indicate  how rapidly that  transition takes place. At $z\sim 3$, when the age of the Universe was $\sim$2.2 Gyr, it has moved substantially towards completion. 
  
 The isotropic solution includes the radial $\alpha_{\parallel}$ component, which in turn incorporates redshift distortion. The transverse component, $\alpha_{\perp}$, provides a determination in both redshift and real space, and is free of redshift distortion.  Comparisons of $h$ and $\Omega_m$ values derived using this measure, as opposed to values derived with the isotropic measure, $\alpha$, show differences of several percent. We conclude that given the clarity of the transverse anisotropic measurements, studies, such as those of  \citet{Aubourg2015,Addison2018,Aghanim2018}, which evaluate cosmological parameters would better reflect reality by omission of the isotropic results.
 
 At the higher redshifts of Ly$\alpha$ studies, $z\simeq$ 2.35, the anisotropic BAO algorithms give decidedly misleading results.  We have replicated an example from \citet{Addison2018}, in which use of the BAO algorithms points to a mean value of $\Omega_m\sim$ 0.197, while the flat $\Lambda$CDM analysis yields a mean value closer to 0.239.  
  
 The next generation of BAO studies, \citep{Vargas2019,Euclid2019,Dore2019}, will improve precision to $\sim$ 0.3 percent.  Under those circumstances, even uncertainties between the fiducial damping parameter, $\Sigma_{nl, \text{fid}}$ and the actual damping parameter, $\Sigma_{nl}$, can introduce errors comparable to that precision.  We have demonstrated that the BAO algorithms are not suitable for that tightened state of affairs.  On the other hand, an analysis assuming a flat, $w=-1$, $\Lambda$CDM cosmology, such as we have done herein, accords a straightforward procedure for preliminary appraisals.
 
\acknowledgments

We acknowledge the use of the Legacy Archive for Microwave Background Data Analysis (LAMBDA), part of the High Energy Astrophysics Science Archive Center (HEASARC).  HEASARC/LAMBDA is a service of the Astrophysics Science Division at the NASA Goddard Space Flight Center.

\appendix
\section{Validity of assumption, \lowercase{$r_{p, \text{fid}}^{*}/r_{p}^{*}$$\simeq \alpha$}}
\label{sec:appen1}
We write,
\begin{equation}\label{E:rast}
\alpha=\frac{r_{p, \text{fid}}}{r_{p}}=\frac{r_{p, \text{fid}}^{*} + \Delta_{\text{fid}}}{r_{p}^{*} +\Delta}.
\end{equation}
Here, $\Delta$ and $\Delta_{\text{fid}}$ represent the shifts introduced by large-scale non-linearities. Let $\alpha=1+\gamma$. The ratio $r_{p, \text{fid}}^{*}/r_{p}^{*}$ is then, 
\begin{equation}
\label{rast2}
\frac{r_{p, \text{fid}}^{*}}{r_{p}^{*}}=\alpha +\frac{ \gamma \Delta + (\Delta-\Delta_{\text{fid}})}{r_{p}^{*}},
\end{equation}
and the approximation is demonstrated, since $\mid$$\gamma$$\mid$ is generally $<$0.05, $\mid$$\Delta$$\mid$/$r_{p}^{*}$ is of the order 0.005--0.01, and the ($\Delta-\Delta_{\text{fid}}$)/$r_{p}^{*}$ term is of the same magnitude as $\gamma \Delta$/$r_{p}^{*}$. 

\section{BAO equations behavior as \lowercase{$z$} $\Rightarrow{0}$}
\label{sec:appen}

As  $z \Rightarrow 0$, the ratios $D_{M}/D_{M, \text{fid}}$ and $D_{V}/D_{V, \text{fid}}$ both $\Rightarrow$ ${h_{\text{fid}}}/h$, the familiar short range measure. Focusing on Equation~\ref{E:DM1}, we rewrite it as,
\begin{equation}\label{E:DM2}
\alpha_{\perp}=\frac{r_{d, \text{fid}}}{r_{d}}\frac{h_{\text{fid}}}{h},
\end{equation}
and derive expressions for each side of the equation.  For $\alpha_{\perp}$, from the various fit line formulas, Equations~\ref{E:h476}, \ref{E:h588}, \ref{E:h833}, and \ref{E:h333}, we obtain,
\begin{equation}\label{E:fit}
\alpha_{\perp}=\frac{h_{\text{fid}}}{h}\frac{(1-C/h_{\text{fid}})}{(1-C/h)}.
\end{equation}
Here C is the constant in the fit formulas, e.g., 0.1334 in Equation~\ref{E:h476}.  We reformulate that expression using the following approximation,
\begin{widetext}
\begin{equation}\label{E:cexpan}
\alpha_{\perp}\simeq \frac{h_{\text{fid}}}{h}\big[1+C/h_{\text{fid}}(1+C/h +C^{2}/h^{2}+C^{3}/h^{3})\frac{\Delta h}{h}\big],
\end{equation}
where $\Delta h=h_{\text{fid}}-h$.

From Equation~\ref{E:rd}, $r_{d, \text{fid}}/r_{d}$ takes the form,
\begin{equation}
\label{E:rd2}
\frac{r_{d, \text{fid}}}{r_{d}}\simeq\frac{(\Omega_{cb}h^{2})^{0.25351}(\Omega_{b}h^{2})^{0.12807}}{(\Omega_{cb, \text{fid}}h_{\text{fid}}^{2})^{0.25351}(\Omega_{b, \text{fid}}h_{\text{fid}}^{2})^{0.12807}},
\end{equation}
 Making use of Equation~\ref{E:degen} we find,
\begin{equation}
\label{E:rd3}
\frac{r_{d, \text{fid}}}{r_{d}}\simeq\frac{(\Omega_{b}h^{2})^{0.12807}}{(\Omega_{b, \text{fid}}h_{\text{fid}}^{2})^{0.12807}} [1 + 0.00168\Delta h]\bigg(\frac{h_{\text{fid}}}{h}\bigg)^{0.25351},
\end{equation}
\end{widetext}
The factor, $ [1 + 0.00168\Delta h]$, stems from subtracting the neutrino mass fraction, $\Omega_{\nu}h^{2}=0.00064$ from $\Omega_{m}h^{2}$ to obtain $\Omega_{cb}h^{2}$.  That factor can be dropped, with little loss of accuracy, leading to the expression,
\begin{equation}
\label{E:rd4}
 \frac{r_{d, \text{fid}}}{r_{d}}\simeq\frac{(\Omega_{b}h^{2})^{0.12807}}{(\Omega_{b, \text{fid}}h_{\text{fid}}^{2})^{0.12807}}\bigg(\frac{h_{\text{fid}}}{h}\bigg)^{0.25351}.
\end{equation}
Within a range of values, such as found in Table~\ref{tab:1}, the ratio, $(\Omega_{b}h^{2})^{0.12807}/(\Omega_{b, \text{fid}}h_{\text{fid}}^{2})^{0.12807}$, differs from unity, at most, by less than three tenths of a per cent, and therefore for this exercise we also drop that term

The right side of Equation~\ref{E:DM2} is now expressed as,
\begin{equation}
\label{E:rhs1}
\frac{r_{d, \text{fid}}}{r_{d}}\frac{h_{\text{fid}}}{h}\simeq\bigg(\frac{h_{\text{fid}}}{h}\bigg)^{1.25351}
\end{equation}
Placing the equation in the same format as Equation~{\ref{E:cexpan}}, we have,
\begin{equation}
\label{E:rhs}
\frac{r_{d, \text{fid}}}{r_{d}}\frac{h_{\text{fid}}}{h}\simeq \frac{h_{\text{fid}}}{h}(1 + 0.25351\frac{\Delta h}{h}),
\end{equation}

Though it is not obvious that Equations~\ref{E:fit}/\ref{E:cexpan} and \ref{E:rhs1}/\ref{E:rhs} give very similar results, a couple of examples should suffice as a demonstration. First consider the test set $h_{\text{fid}}=0.66$, $h=0.6732$, and Equation~\ref{E:h476}, $\Sigma_{nl}=4.76$ $h^{-1}$ Mpc, C=0.1334. For those values, Equation~\ref{E:fit} gives $\alpha_{\perp}=0.9755$, while Equation~ \ref{E:rhs1}  also yields $r_{d, \text{fid}}h_{\text{fid}}/(r_{d}h)=0.9755$.  At the other extreme of the range for $h$, and taking a value of $\alpha_{\perp}$ further removed from unity, we select from the test sets, $h_{\text{fid}}=0.6987$, and $h=0.7344$, with $\Sigma_{nl}=5.88$ $h^{-1}$ Mpc, C=0.1351.  For these conditions we find, $\alpha_{\perp}=0.9404$, and  $r_{d, \text{fid}}h_{\text{fid}}/(r_{d}h)=0.9394$.  For $\Sigma_{nl}=8.33$ $h^{-1}$ Mpc, $\alpha_{\perp}=0.9383$.  Thus, throughout the range of relevant values for $h$ and $\Sigma_{nl}$,  $r_{d, \text{fid}}h_{\text{fid}})/(r_{d}h)$ and $\alpha_{\perp}$ converge towards a common value.  

This demonstration is not satisfactory from the perspective of lacking a clear analytical limit.  That problem stems from the empirical nature of both Equation~\ref{E:rd}, defining $r_{d}$, and of the fit lines that define $\alpha_{\perp}$.  We cannot overcome that shortcoming in this presentation, but the approximations of Equations~{\ref{E:cexpan}} and {\ref{E:rhs}} perhaps make the demonstration more palatable. They take the same form, the difference being in the terms multiplying $\Delta h/h$. The demonstration works because those factors are small, affecting the third and fourth significant figures. As a result, some degree of inequality can be tolerated, while still bringing about convergence.

\end{document}